# Are Security Cues Static? Rethinking Warning and Trust Indicators for Life Transitions

Sarah Tabassum

University of North Carolina at Charlotte, stabass2@charlotte.edu

Security cues, such as warnings and trust signals, are designed as stable interface elements, even though people's lives, contexts, and vulnerabilities change over time. Life transitions including migration, aging, or shifts in institutional environments reshape how risk and trust are understood and acted upon. Yet current systems rarely adapt their security cues to these changing conditions, placing the burden of interpretation on users. In this Works-in-Progress paper, we argue that the static nature of security cues represents a design mismatch with transitional human lives. We draw on prior empirical insights from work on educational migration as a motivating case, and extend the discussion to other life transitions. Building on these insights, we introduce the Transition-Aware Security Cues (TASeC) framework and present speculative design concepts illustrating how security cues might evolve across transition stages. We invite HCI to rethink security cues as longitudinal, life-centered design elements collectively.

## 1 INTRODUCTION

In everyday life, people increasingly rely on digital devices to carry out routine and consequential tasks, from communication and finance to work and civic participation. Within these systems, security cues such as warnings, indicators, trust signals, and prompts serve as key interface mechanisms for communicating risk, safety, and trustworthiness to users [4, 11]. Browser warnings, permission dialogs, padlock icons, and confirmation messages shape how people decide what to click, what to share, and whom to trust [1, 10]. Much HCI and usable security research has therefore focused on improving the clarity, visibility, and effectiveness of these cues to support safer decision-making. At the same time, people's lives are rarely stable. Many experience periods of change such as migration or displacement, aging, illness or recovery, entering new institutions, or shifts in digital literacy. During these life transitions, users' knowledge, confidence, and vulnerability can change rapidly, altering what feels safe, risky, or trustworthy. Yet security cues are typically designed to remain constant, regardless of these changes.

This creates a mismatch between static interface designs and dynamic human lives. Current security cues implicitly assume stable users, stable contexts, and stable interpretations, placing the burden of sensemaking on users precisely when they may be most uncertain. For people navigating life transitions, this assumption can make security cues difficult to interpret, easy to ignore, or misleading. In this paper, we argue that security cues are designed as if lives were static, while lives are fundamentally transitional. To address this gap, we propose a design framework for transition-aware security cues that treats life transitions as first-class design conditions rather than edge cases. The framework reframes warnings, indicators, and trust signals as longitudinal, life-centered design elements that evolve as users' familiarity and confidence change. We illustrate this approach through speculative design concepts that demonstrate how cue strategies can shift across different phases of a transition.

## 2 BACKGROUND AND PROBLEM FRAMING: SECURITY CUES IN TRANSITIONAL LIVES

Security cues and prompts are widely used interface mechanisms for communicating risk, trust, and safety to users [4, 6]. Examples include browser warnings about insecure websites, permission dialogs requesting access to sensitive data, and visual indicators such as padlock icons. Prior HCI and usable security research has examined how these cues influence user understanding, trust judgments, and decision-making, often focusing on improving clarity, visibility, and perceived control [1, 11]. At their core, security cues function as forms of risk communication that guide user behaviour in situations involving uncertainty or potential harm.

Despite their central role, most security cues are designed around an implicit assumption of stability. The same cues are typically presented to all users, in the same form, and are expected to carry consistent meaning over time. Research on trust indicators and phishing shows that users interpret cues such as logos, icons, or visual legitimacy markers differently depending on experience and context, yet systems rarely account for this variability [4]. Similarly, studies of permission interfaces demonstrate that even when explanations and visual cues are provided, users may feel overwhelmed or misunderstand their implications, challenging the assumption that more information always leads to better decisions [6]. These designs presume stable knowledge, stable context, and stable users, an assumption that becomes fragile in real-world use.

Life transitions make this fragility especially visible and can be understood as a design stress test for security cues. Transitions such as migration, displacement, aging, illness, or becoming digitally connected often involve shifts in knowledge, confidence, power, and trust relationships [3, 7, 8]. During these periods, users may rely more heavily on institutions, appearances of legitimacy, or familiar patterns, while also facing new forms of risk. However, security cues typically remain static, placing the burden of interpretation entirely on users as their contexts change. This creates a core mismatch: lives change faster than interfaces. As interpretation contexts shift while cues remain fixed, their meaning and effectiveness drift, motivating the need for design approaches that treat life transitions as central conditions for security cue design rather than as edge cases.

## 3 A FRAMEWORK FOR TRANSITION-AWARE SECURITY CUES (TASEC)

To address the mismatch between static security cues and dynamic human lives, we propose the *Transition-Aware Security Cues (TASeC)* framework. TASeC is a design-oriented framework for reasoning about how security warnings, indicators, and trust signals should evolve as people move through life transitions. Rather than treating cues as universal and persistent interface elements, TASeC frames them as longitudinal, life-centered components whose form, prominence, and function change with users' contexts. Life transitions such as migration [8], displacement, aging [5], or shifts in digital literacy [2] involve changing familiarity, confidence, and vulnerability, causing the same cue to carry different meanings over time. TASeC links life phase, user state, and cue strategy to support more appropriate security communication across these changes. The framework is conceptual and generative, intended to support design reasoning rather than prescribe or validate a specific system.

### 3.1 Design Principles

TASeC is guided by four design principles. **Temporality** emphasizes that security cues should change over time rather than remain fixed. **Scaffolding** supports providing more explanation and guidance during periods of uncertainty, with reduced support as confidence grows. **Reversibility** focuses on adjusting friction and guidance to allow users to pause, reassess, or redirect actions safely, rather than relying on irreversible warnings. **Calibration** highlights that cues should match users' current confidence and capability, recognizing that the same cue may be helpful in one context and intrusive or insufficient in another.

### 3.2 Framework Model

TASeC organizes security cue design around three dimensions: *life phase, user state,* and *cue strategy* (Figure 1)

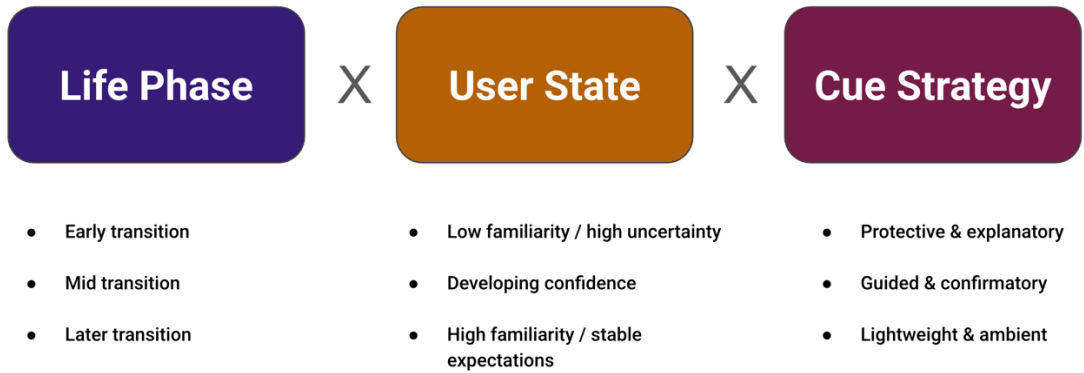

Figure 1: The figure illustrates how different cue strategies can be selected based on changing life contexts and user confidence, rather than as a fixed or deterministic flow.

Life phase captures a person's position within a transition, user state reflects situated knowledge and confidence, and cue strategy describes how risk and trust are communicated. The framework encourages selecting different cue strategies across early, mid, and later phases, making explicit a design space for adapting security cues across time and context.

## 4 ILLUSTRATIVE DESIGN CONCEPTS FOR THE TASEC FRAMEWORK

We illustrate the TASeC framework through a concrete and familiar high-risk scenario: receiving a potentially fraudulent SMS message. This example demonstrates how security cues can adapt across life transition stages rather than remaining static.

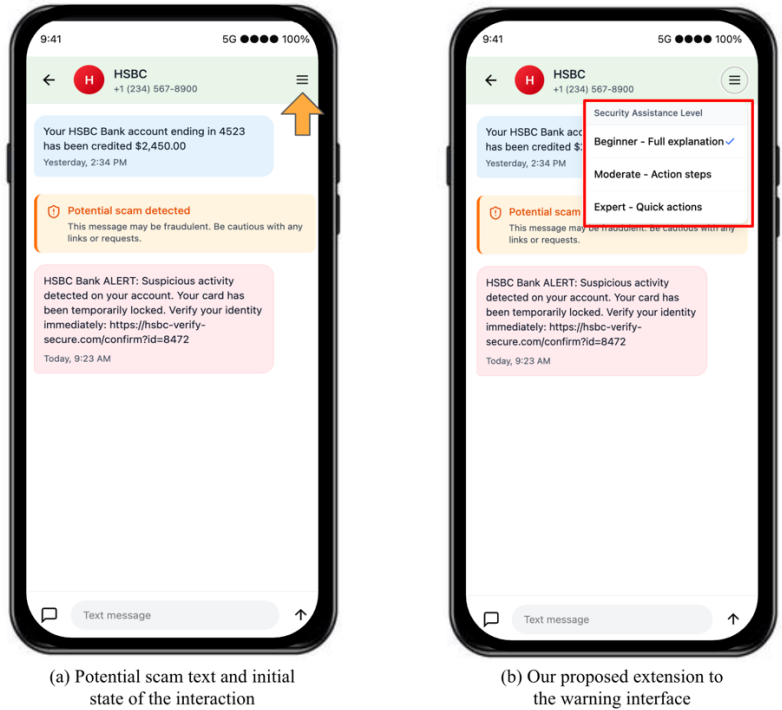

(a) Potential scam text and initial state of the interaction

(b) Our proposed extension to the warning interface

Figure 2: Initial state showing a standard potential scam warning for an incoming SMS message

Figure 2 (a) shows a standard potential scam warning commonly used on mobile platforms. Such warnings typically appear as brief alerts that signal risk but provide little contextual explanation. While effective in some cases, they assume that all users interpret and respond to the cue in the same way. To address this limitation, we propose an extension to the warning interface. As shown in Figure 2 (b), selecting the hamburger menu reveals three support options aligned with different transition stages: **Beginner** (full explanation), **Moderate** (action steps with confirmation), and **Expert** (quick actions with minimal interruption).

## 4.1 Beginner: Full Explanation

The Beginner option supports early transition stages marked by low familiarity and high uncertainty. As shown in Figure 3 (a), this interface provides three components: an explanation of why the message appears suspicious, clear step-by-step guidance on what to do next, and an option to report the message. This mode introduces more explanation and friction to help users pause and make informed decisions.

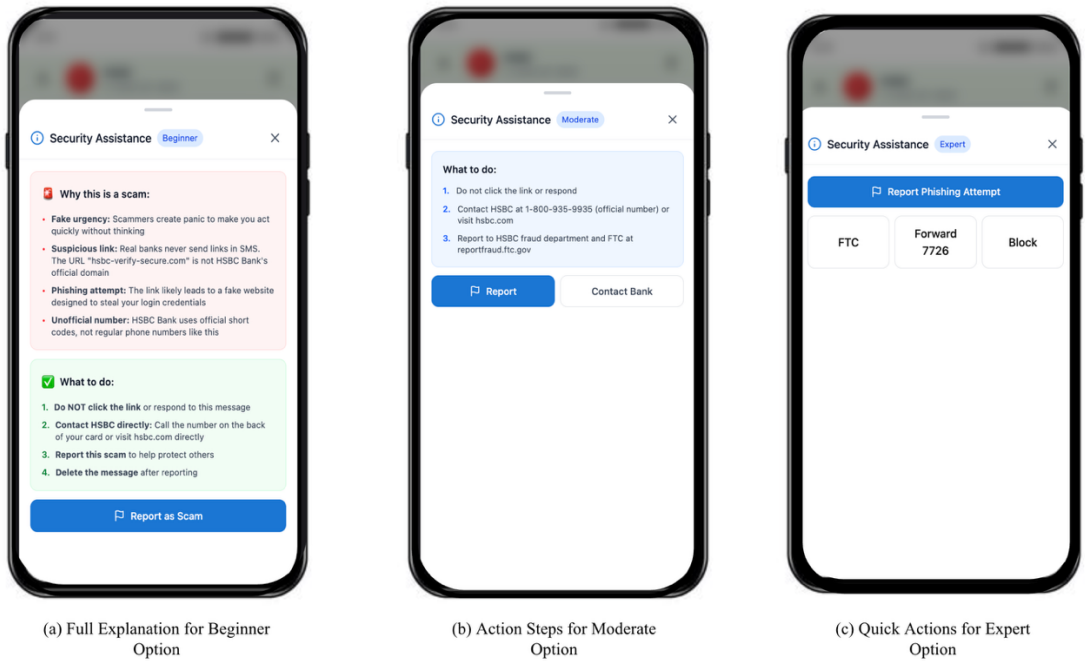

Figure 3: Three TASeC interaction options: (a) Beginner with full explanation, (b) Moderate with action steps, and (c) Expert with quick actions.

## 4.2 Moderate: Action Steps

The Moderate option targets mid transition stages, where users recognize common scam cues but benefit from confirmation. As shown in Figure 3 (b), this interface emphasizes concise action steps, retains reporting functionality, and offers the option to contact a trusted institution for verification. This mode balances guidance with growing user confidence.

## 4.3 Expert: Quick Action Steps

The Expert option supports later transition stages, where users are confident and familiar with scam patterns. As shown in Figure 3(c), the interface minimizes interruption and offers fast actions such as reporting, blocking the sender, or forwarding the message to a reporting service. This mode prioritizes low friction, autonomy, and subtle security cues.

## 5 REFLECTIONS AND IMPLICATIONS

The transition-aware approach proposed in this paper raises some open questions and design tensions that warrant careful consideration. First, a central challenge concerns how life transitions are identified. Detecting transitions automatically risks misclassification and raises concerns about inference and surveillance, while relying entirely on user self-identification may increase burden or exclude those who are unaware of available support. Future work is needed to explore how transition-aware security cues can be activated in ways that are transparent, optional, and respectful of user autonomy. A second tension involves the balance between adaptation and control. While adaptive cues can provide valuable support during periods of vulnerability, over-adaptation risks reducing user agency or creating dependency on system guidance.

Designers must therefore consider when adaptation is beneficial and when it may become intrusive or counterproductive. Also, not all situations call for adaptation. Some users may prefer stable and predictable interfaces, even during periods of change, highlighting the importance of reversibility and user choice.

Beyond these challenges, the framework proposed in this paper has broader implications for HCI and usable security research. In usable security and trust and safety design, it invites a shift away from one-size-fits-all warnings toward designs that account for changing user states and lived experience over time [8, 9]. More broadly for HCI, it contributes to ongoing discussions about designing not just for isolated interactions, but for processes that unfold across time, life events, and shifting contexts. The key shift suggested by this work is a move from designing security cues as static interface elements to designing them as **life-centered processes** that evolve alongside users. Treating life transitions as first-class design conditions opens new opportunities for rethinking how systems communicate risk and support decision-making in moments of uncertainty, change, and learning.

## 6 CONCLUSION

Security warnings, indicators, and trust signals are typically designed as static and universal, despite the fact that people's lives, contexts, and vulnerabilities change over time. In this paper, we argued that this mismatch becomes particularly visible during life transitions, when users must interpret security cues under uncertainty and shifting confidence. To address this gap, we introduced the *Transition-Aware Security Cues (TASeC)* framework, which reframes security cues as longitudinal, life-centered design elements rather than fixed interface components. By linking life phase, user state, and cue strategy, TASeC opens a design space for developing adaptive and supportive security cues that better align with changing lived experience aka the context of the users. We hope this framework encourages future work in usable security, trust and safety, and HCI to treat life transitions not as edge cases, but as central conditions for designing how systems communicate risk and trust.